\documentclass{webofc}
\usepackage[varg]{txfonts}   
\begin{document}
\title{The LbL contribution to the muon g-2 from the axial-vector mesons exchanges within the nonlocal quark model}
%
%

\author{
 \firstname{A.E.} \lastname{Dorokhov}\inst{1,2}
 \and
  \firstname{A.P.} \lastname{Martynenko}\inst{2}
  \and
  \firstname{F.A.} \lastname{Martynenko}\inst{2}
  \and
  \firstname{A.E.} \lastname{Radzhabov}\inst{2,3}\fnsep\thanks{\email{aradzh@icc.ru}} \and
  \firstname{A.S. } \lastname{Zhevlakov}\inst{3}
}

\institute{Joint Institute of Nuclear Research, BLTP,141980, Moscow region, Dubna, Russia
\and Samara University, 443086, Samara, Russia
\and Matrosov Institute for System Dynamics and Control Theory SB RAS, 664033, Irkutsk, Russia
          }

\abstract{%
The light-by-light contribution from the axial-vector (AV) mesons exchanges to the muon anomalous magnetic moment is estimated in the framework of the nonlocal chiral quark model. The preliminary answer for contributions from $a_1$ and $f_1$ mesons to $(g-2)_\mu$ is $0.34\cdot10^{-11}$ and does not support the Melnikov-Vainshtein estimate $2.2\cdot10^{-11}$.
}
\maketitle
%

The light-by-light contributions to the anomalous magnetic moment of muon of the dynamical quark loop, i.e. contact type, and of the light pseudoscalar and scalar resonances exchanges within the nonlocal chiral quark model was calculated in our previous works \cite{Dorokhov:2011zf,Dorokhov:2012qa,Dorokhov:2015psa}.
In the present work we extend our model calculations by including the vector--axial-vector sector. To this aim 
the model parameters should be refitted to the pion observables, the mixing of the pseudoscalar and longitudinal part of the axial-vector mesons as well as to the $\rho-\gamma$ mixing. 

It is interesting to note that in \cite{Dorokhov:2017nzk} is shown that the axial-vector exchange interaction in muonic hydrogen gives essential contribution to hyperfine splitting. For hyperfine splitting the axial-vector contribution is even bigger than the pion one \cite{Dorokhov:2017gst}.

For these purposes we consider the $SU(2)\times SU(2)$ chiral quark model where the pseudoscalar--scalar and vector--axial-vector sectors are included.
The corresponding Lagrangian 
is 
\begin{eqnarray}
&&\mathcal{L}= \mathcal{L}_{free}+\mathcal{L}_{PS,S}+\mathcal{L}_{V,AV} 
 ,\quad 
\mathcal{L}_{free} = \bar{q}(x)(i \hat{\partial}-m_c)q(x), \\
&&
\mathcal{L}_{PS,S} = \frac{G_1}{2}[J_S^a(x)J_S^a(x)+J^a_P(x)J^a_P(x)] ,\quad 
\mathcal{L}_{V,AV} = \frac{G_2}{2}[J_V^a(x)J_V^a(x)+J^a_{AV}(x)J^a_{AV}(x)],  \nonumber
\end{eqnarray}
where $m_c$ is the current quark mass matrix with diagonal elements $m_c^u=m_c^d$, $G_1$ and $G_2$ are the coupling constants in pseudoscalar--scalar and vector--axial-vector sectors, respectively. 
The nonlocal quark currents are given by
\begin{align}
J_{M}^{a}(x)=\int d^{4}x_{1}d^{4}x_{2}\,f(x_{1})f(x_{2})\, \bar{q}
(x-x_{1})\,\Gamma_{M}^{a}q(x+x_{2}),\label{eq2}
\end{align}
with $M=S,P,V,AV$ and $\Gamma_{{S}}^{a}=\lambda^{a}$, $\Gamma_{{P}}^{a}=i\gamma^{5}\lambda^{a}$, $\Gamma_{{V}}^{a}=\gamma^{\mu}\lambda^{a}$, $\Gamma_{{AV}}^{a}=\gamma^{5}\gamma^{\mu}\lambda^{a}$. For the $SU(2)$ model,
the flavour matrices are $\lambda^{a}\equiv\tau^{a}$, $a=0,..,3$ with $\tau^0=1$.
In Eq. (\ref{eq2}), $f(x)$ is the form factor reflecting the nonlocal properties of the QCD vacuum. The action of the model can be bosonized by the usual Hubbard-Stratonovich trick with introduction of auxiliary mesonic fields. 
In order to obtain a physical scalar field with zero vacuum expectation value, the shift of the scalar isoscalar field should be performed
leading to appearance of the dynamical momentum-dependent quark mass 
\begin{eqnarray}
m(p)=m_c+m_{dyn}f^2(p),\quad m_{dyn}= G_1 \frac{8  N_c }{(2 \pi)^4} \int d^4_Ek
\frac{f^2(k)m(k)}{k^2+m^2(k)}.
\end{eqnarray}
The meson vertices without $\pi-a_1$ mixing in momentum space are
\begin{eqnarray}
V_M=g_{M}(k) \Gamma_M f(p_-) f(p_+),
\end{eqnarray}
where $g_{M}(k)$ is the meson renormalization constant, $p_\pm,k$ are the quark and meson momenta, respectively.
The presence of the longitudinal component of axial-vector mesons  leads to modification of the pion vertex as
\begin{eqnarray}
V_\pi=g_{\pi}(k) i\gamma^{5}\lambda^{a}\left(1-\hat{k}\tilde{g}_{\pi}(k)\right) f(p_-) f(p_+).
\end{eqnarray}
Due to the nonlocal interaction, there appear additional contributions to the antiquark-quark-photon vertex and the meson-antiquark-quark-photon vertex which takes the form \footnote{The antiquark-quark-photon and meson-antiquark-quark vertices with two photons appearing due to the nonlocal interaction give contribution only to the scalar-meson transition form factor and are absent for pseudoscalar and axial-vector mesons. }
\begin{align}
&\mathrm{\Gamma}^\mu_{p_2,p_1}=Q (\gamma_\mu-(p_{1}+p_{2})_{\mu}\mathrm{m}^{(1)}(p_1,p_2)), \nonumber\\
&\mathrm{\Gamma}^{M;\mu}_{p_2,p_1,q}= -g_{M}(k)\left(\mathrm{f}^{(1)}(p_{1},p_{1}+q) f(p_{2})(2p_{1}+q)_{\mu}Q\Gamma _{M}+\right.\nonumber\\
&\quad\quad\left.+\mathrm{f}^{(1)}(p_{2},p_{2}-q)f(p_{1})(2p_{2}-q)_{\mu}\Gamma _{M}Q\right) , \label{GammaMppq}
\end{align}
where $\mathrm{m}^{(1)}(p_1,p_2) =(m(p_1^{2}) - m(p_2^{2}))/({p_{1}^2-p_{2}^2})$ is the first order finite-difference.
In the presence of the vector sector, an additional dressing of interaction vertices with photons arises due to $\rho(\omega)\rightarrow \gamma$ transition \cite{Plant:1997jr,Dorokhov:2003kf}. The model parameters are taken from \cite{Plant:1997jr}.


The general form of the axial-vector meson to two-photon transition form factor is \cite{Rosenberg:1962pp,Adler:1969gk}
\begin{align}
&T^{\mu\nu}_\alpha =ie^2\varepsilon_{\rho \sigma \tau \alpha} \biggl\{ A_1q_1^{\tau}g^{\mu \rho}g^{\sigma \nu}
+A_2 q_2^{\tau}g^{\mu \rho}g^{\sigma \nu}
+A_3 q_1^{\nu} q_1^{\rho} q_2^{\sigma}g^{\tau \mu} +\nonumber\\
&\quad\quad\quad\quad
+A_4 q_2^{\nu} q_1^{\rho} q_2^{\sigma}g^{\tau \mu}
+A_5 q_1^{\mu} q_1^{\rho} q_2^{\sigma}g^{\tau \nu}
+A_6 q_2^{\mu} q_1^{\rho} q_2^{\sigma}g^{\tau \nu}  \biggr\}, 
\end{align}
where $p$, $q_1$ and  $q_2$ are momenta of AV meson and photons with indices $\alpha,\mu,\nu$,
$A_i \equiv \mathcal{A}_i(p^2,q_1^2,q_2^2)$. The gauge invariance provides the relations
\begin{eqnarray}
A_2 = q_1^2A_5 + (q_1\cdot q_2)A_6, \quad
A_1 = (q_1\cdot q_2)A_3 + q_2^2 A_4
\end{eqnarray}
and the Bose symmetry  leads to%
$\mathcal{A}_1(p^2,q_1^2,q_2^2)=-\mathcal{A}_2(p^2,q_2^2,q_1^2)$, 
$\mathcal{A}_3(p^2,q_1^2,q_2^2)=-\mathcal{A}_6(p^2,q_2^2,q_1^2)$, 
$\mathcal{A}_4(p^2,q_1^2,q_2^2)=-\mathcal{A}_5(p^2,q_2^2,q_1^2)$. 
Alternatively, one can rewrite the amplitude as
\begin{align}
&T^{\mu\nu}_\alpha =ie^2 \varepsilon_{\rho \sigma \tau \alpha} \biggl\{
R^{\mu \rho}_{ q_1, q_2} R^{\nu \sigma}_{ q_1, q_2} \,
(q_1 - q_2)^\tau \, \frac{(q_1 \cdot q_2)}{M_A^2} \, F^{(0)}_{AV\gamma^\ast\gamma^\ast}(p^2,q_1^2, q_2^2)
\nonumber \\
&\quad\quad\quad + \, R^{\nu \rho}_{ q_1, q_2} Q_1^\mu
q_1^\sigma \, q_2^\tau \,  \frac{1}{M_A^2} \, F_{AV\gamma^\ast\gamma^\ast}^{(1)}(p^2,q_1^2, q_2^2) 
 +R^{\mu \rho}_{ q_1, q_2} Q_2^\nu
q_2^\sigma \, q_1^\tau \, \frac{1}{M_A^2} \, F^{(1)}_{AV\gamma^\ast\gamma^\ast}(p^2,q_2^2, q_1^2) \biggr\}
, \nonumber \\
&\quad R^{\mu \nu} _{ q_1, q_2}  = - g^{\mu \nu} + \frac{1}{X} \,
\bigl \{ (q_1 \cdot q_2)\left( q_1^\mu \, q_2^\nu + q_2^\mu \, q_1^\nu \right)
- q_1^2 \, q_2^\mu \, q_2^\nu  - q_2^2 \, q_1^\mu \, q_1^\nu \bigr\}
,  \\
&\quad \quad Q_1^\mu=
q_1^\mu -q_2^{\mu}  \frac{q_1^2}{(q_1 \cdot q_2)} 
,\quad
Q_2^\nu=
q_2^\nu - q_1^{\nu} \frac{q_2^2}{(q_1 \cdot q_2)}
,\quad X = (q_1 \cdot q_2)^2 - q_1^2 q_2^2,\nonumber
\end{align}
where $R^{\mu \nu} _{ q_1, q_2}$  is the totally transverse tensor,  $Q_1^\mu$ and $Q_2^\nu$ are transverse with respect to $q_1$ and $q_2$, respectively.

According to the Landau--Yang theorem, the axial-vector mesons can not decay into two real photons. However, the coupling of $1^{++}$ mesons to two photons is allowed in the case when one or both photons are virtual. The two-photon ``decay'' width is defined as
\begin{align}
&\tilde\Gamma_{\gamma^\ast\gamma^\ast}(AV) = \lim_{Q^2\rightarrow 0} \frac{M_A^2}{Q^2} \Gamma^{\mathrm{TS}}_{\gamma\gamma^\ast} =\frac{\pi\alpha^2M^5_{A}}{12}[F^{(1)}_{AV\gamma^\ast\gamma^\ast}(M_{A}^2,0,0)]^2. 
\end{align}
\begin{figure}[t]
	\centering
	\includegraphics[width=0.6\textwidth]{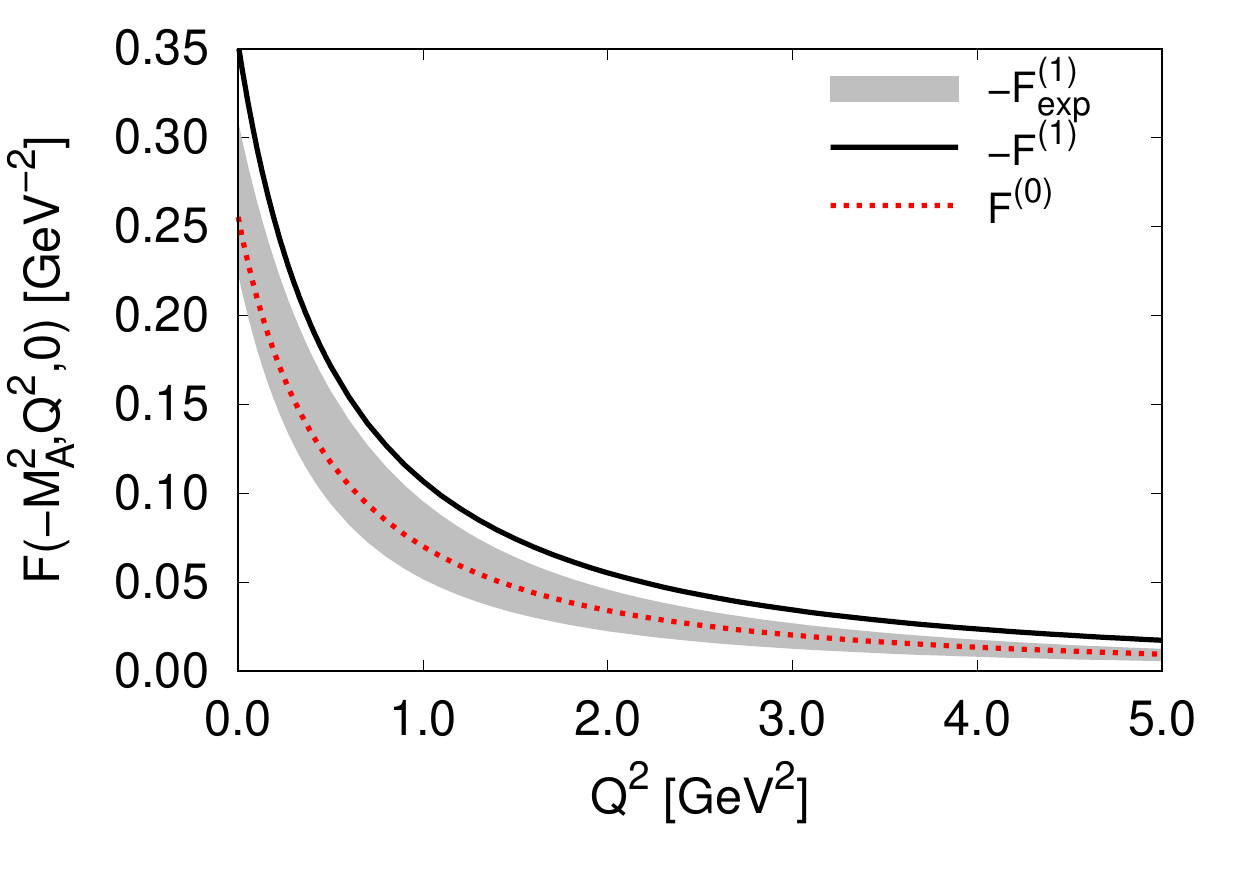}
	\caption{Behaviour of the form-factors for on mass-shell $f_1(1285)$. Nonlocal model calculation: black solid line is  $-F^{(1)}_{AV\gamma^\ast\gamma^\ast}(M_{A}^2,Q^2,0)$, red dotted is $F^{(0)}_{AV\gamma^\ast\gamma^\ast}(M_{A}^2,Q^2,0)$. Shaded region is the result of L3 collaboration for $-F^{(1)}_{AV\gamma^\ast\gamma^\ast}(M_{A}^2,Q^2,0)$. }
	\label{fig-1}       
\end{figure}

\begin{figure}[t]
	\centering
\begin{center}
	\begin{tabular*}{\columnwidth}{@{}ccc@{}}
		\raisebox{-0.5\height}{\resizebox{0.2\textwidth}{!}{\includegraphics{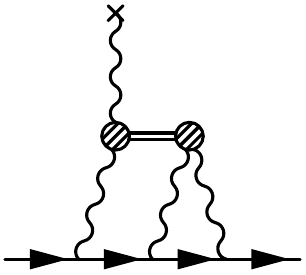}}}&
		\raisebox{-0.5\height}{\resizebox{0.2\textwidth}{!}{\includegraphics{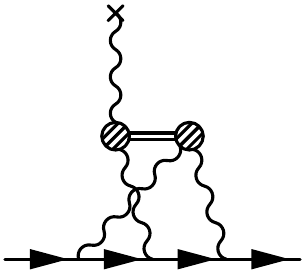}}}&
		\raisebox{-0.5\height}{\resizebox{0.2\textwidth}{!}{\includegraphics{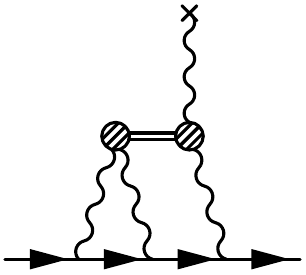}}}
	\end{tabular*}
\end{center}	
	\caption{LbL contribution from intermediate meson exchanges.}
	\label{fig:LbL}
\end{figure}

At present, we have only few experimental data on the $1^{++}$ meson transition form factor into two photons.
The L3 Collaboration studied the reaction
$e^+e^- \to e^+e^-\gamma^\ast\gamma^\ast \to e^+e^-f_1(1285)\to e^+e^- \eta \pi^+\pi^-$
and the $f_1(1285)$ meson transition form factor for the case when one of the photons
is real and another one is virtual was extracted \cite{Achard:2001uu}. The comparison of the model calculations for the axial-vector meson form factors with the results of the L3 collaboration is shown in fig.\ref{fig-1}. One can see that for on-mass shell meson, the agreement of the model calculation with the experiment is reasonable.
The muon anomalous magnetic moment is defined from the following projection \cite{Brodsky:1966mv}
\begin{align}
a_{\mu }^{\mathrm{HLbL}}&=\frac{1}{48m_{\mu }}\mathrm{Tr}\left( (\hat{p}%
+m_{\mu })[\gamma ^{\rho },\gamma ^{\sigma }](\hat{p}+m_{\mu })\mathrm{\Pi }%
_{\rho \sigma }(p,p)\right) , \nonumber \\
& \mathrm{\Pi }_{\rho \sigma }(p^{\prime },p)=-ie^{6}\int \frac{d^{4}q_{1}}{%
	(2\pi )^{4}}\int \frac{d^{4}q_{2}}{(2\pi )^{4}}\frac{1}{%
	q_{1}^{2}q_{2}^{2}(q_{1}+q_{2}-k)^{2}}\times  \label{P4gamProject}\\
& \quad  \times \gamma ^{\mu }\frac{\hat{p}^{\prime }-\hat{q}%
	_{1}+m_{\mu }}{(p^{\prime }-q_{1})^{2}-m_{\mu }^{2}}\gamma ^{\nu }\frac{\hat{%
		p}-\hat{q}_{1}-\hat{q}_{2}+m_{\mu }}{(p-q_{1}-q_{2})^{2}-m_{\mu }^{2}}\gamma
^{\lambda }
\frac{\partial }{\partial k^{\rho }}\mathrm{\Pi }_{\mu \nu
	\lambda \sigma }(q_{1},q_{2},k-q_{1}-q_{2}),  \nonumber
\end{align}%
where $m_{\mu }$ is the muon mass, 
the static limit $k_{\mu }\equiv (p^{\prime }-p)_{\mu }\rightarrow 0$ is implied. Four-rank polarization tensor $\mathrm{\Pi }_{\mu \nu\lambda \sigma }$ is saturated by resonances similarly to the pseudoscalar (scalar) case, see fig. \ref{fig:LbL}. 
Then, by averaging over the direction of muon momentum the result for $a_{\mu }^{\mathrm{HLbL}}$ becomes a three-dimensional integral with 
the radial variables of integration $Q_{1},Q_{2}$ and the angular variable \cite{Dorokhov:2012qa,Jegerlehner:2009ry,Jegerlehner:2017gek}.
After integration over the angular variable the LbL contribution can be represented in the form \cite{Dorokhov:2015psa}
\begin{equation}
a_{\mu}^{\mathrm{LbL}}=\int\limits_{0}^{\infty}dQ_{1}\int\limits_{0}^{\infty
}dQ_{2}\,\,\rho(Q_{1},Q_{2}), \label{aLbL4}%
\end{equation}
where $\rho(Q_{1},Q_{2})$ is density.

\begin{table}
	\centering
	\caption{Contribution to (g-2) of muon from axial-vector meson exchanges. Most of contribution comes from $f_1$.}
	\label{tab-1}       
	\begin{tabular}{|c|c|c|c|c|c|c|}   \hline
	model & AV contribution in $10^{-10}$&\\
	\hline
	ENJL (BPP) & $0.25 \pm 0.1$ &   \cite{Bijnens:1995cc} \\
	HLS (HKS) & $0.2 \pm 0.1$  &  \cite{Hayakawa:1995ps}  \\
	LMD (MV)& $2.2\pm 0.5$  & \cite{Melnikov:2003xd} \\
	LMD (PdRV)& $0.15\pm 0.1$& \cite{Prades:2009tw} \\
	LMD (PV) & $0.64 \pm 0.2$ &\cite{Pauk:2014rta} \\
	LMD (J)& $0.755 \pm 0.271$ &\cite{Jegerlehner:2017gek} \\
	\hline
	This work & $0.67 $, $0.34$ &\\
	\hline
\end{tabular}
\end{table}

In the nonlocal chiral quark model the separate result for contribution of $a_1(1260)$ and $f_1(1285)$ with full kinematic dependence 
is $0.67 \cdot 10^{-11}$. However, due to decrease of the pion contribution resulting from the $\rho-\gamma$ and $\pi-a_1$ mixings, the axial-vector contribution becomes $0.34 \cdot 10^{-11}$. Therefore we present these two numbers in table \ref{tab-1}  for comparison with predictions of other approaches. One can see that our result is visible but not so big as the Melnikov-Vainshtein estimations \cite{Melnikov:2003xd}. 
 
 \begin{figure}[t]
 	\centering
 	\includegraphics[width=0.6\textwidth]{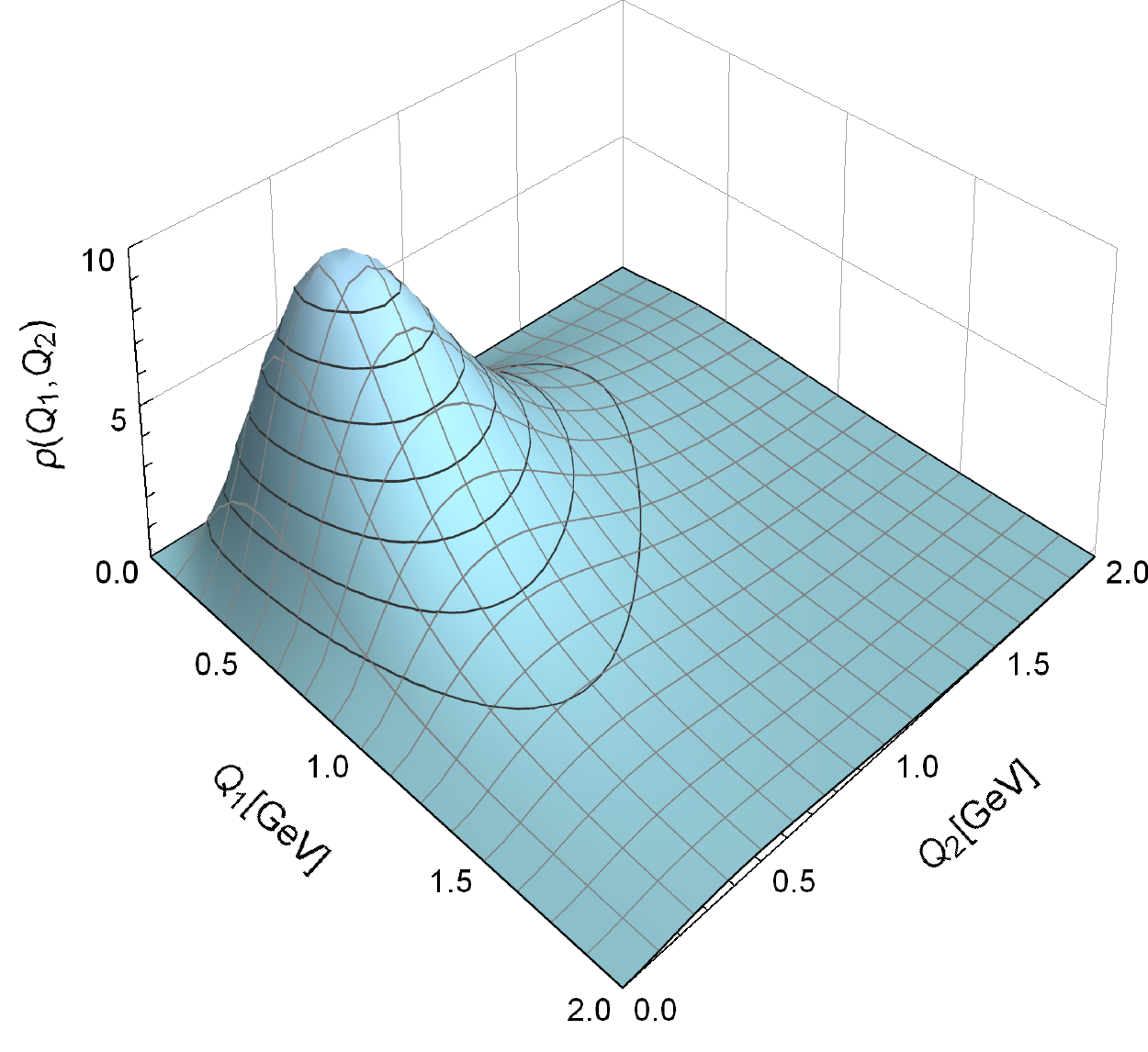}
 	\caption{Behaviour of the density \eqref{aLbL4} in units $10^{-11}$ GeV$^-2$, for the axial-vector meson $f_1$(1285). }
 	\label{fig-density}       
 \end{figure}

The density  $\rho(Q_{1},Q_{2})$  is shown in fig.\ref{fig-density} for the axial-vector meson $f_1$(1285). One can see that peak of density is located at $Q_1,Q_2\sim 0.5$ GeV and most of contribution is in energy domain below $1.5$ GeV. 
 
In future we plan extend our calculations to the sector with strange particles and reestimate the influence of vector--axial-vector sector on the contact term (quark loop). Actually, the presence of the contact term is a main difference between models with quark degrees of freedom and pure mesonic one or the dispersive approach. Up to now, it is not clear how to relate these calculations since the mesonic contributions exist in both approaches, while the contact term is only attributed to the quark models. This contribution is small only for some models \cite{Bijnens:1995cc}, while in calculation in the nonlocal model \cite{Dorokhov:2015psa}, the DSE/BSE approach \cite{Goecke:2010if}, the model with quark box \cite{Pivovarov:2001mw}, C$_\chi$QM \cite{Greynat:2012ww} this contribution is not small and even bigger than resonance one. Important to note, that in the nonlocal chiral quark model just the contact term guarantees the correct QCD asymptotics \cite{Dorokhov:2003kf,Dorokhov:2011zf}.

The work is supported by Russian Science Foundation (grant No. RSF 18-12-00128).

\end{document}